\documentclass[usenatbib,onecolumn]{mn2e}
\usepackage{graphicx}
\usepackage{amsmath}
\usepackage{tabularx}
\usepackage{inputenc}
\usepackage{float}
\usepackage{longtable}
\graphicspath{{images/}}
\newcommand{\aj}{{AJ}}
\newcommand{\apj}{{ApJ}}
\newcommand{\apjs}{{ApJS}}
\newcommand{\aap}{{A\&A}}
\newcommand{\mnras}{{MNRAS}}

\newcommand{\actaa}{{AcA}}

\def\aj{AJ}%
\def\actaa{Acta Astron.}%
\def\apj{ApJ}%
\def\apjs{ApJS}%
\def\apss{Ap\&SS}%
\def\aap{A\&A}%
\def\mnras{MNRAS}%
\def\pasp{PASP}%
\def\pasj{PASJ}%
\title[Binary with sudden and continuous period changes]{V752 Cen - A triple-lined spectroscopic contact binary with sudden and continuous period changes}

\author[X. Zhou et. al]{X. Zhou$^{1,2,3,4}$ \thanks{E-mail: zhouxiaophy@ynao.ac.cn}, B. Soonthornthum$^{2}$, S.-B. Qian$^{1,3,4,5}$, E. Fern\'{a}ndez, Laj\'{u}s$^{6,7}$
\\
$^{1}$ Yunnan Observatories, Chinese Academy of Sciences (CAS), P.O. Box 110, 650216 Kunming, China\\
$^{2}$ National Astronomical Research Institute of Thailand, 260 Moo 4, T. Donkaew,  A. Maerim, Chiangmai, 50180, Thailand\\
$^{3}$ Key Laboratory of the Structure and Evolution of Celestial Objects, Chinese Academy of Sciences, P. O. Box 110, 650216 Kunming, China\\
$^{4}$ Center for Astronomical Mega-Science, Chinese Academy of Sciences, 20A Datun Road, Chaoyang District, Beijing, 100012, China\\
$^{5}$ University of the Chinese Academy of Sciences, Yuquan Road 19\#, Sijingshang Block, 100049 Beijing, P. R. China\\
$^{6}$ Facultad de Ciencias Astron\'{o}micas y Geof\'{i}sicas, Universidad Nacional de La Plata, Paseo del Bosque s/n, 1900, La Plata, Pcia. Bs. As., \\ Argentina\\
$^{7}$ Instituto de Astrof\'isica de La Plata (CCT La plata - CONICET/UNLP), Argentina\\
}

\begin{document}

\date{\today}

\pagerange{\pageref{firstpage}--\pageref{lastpage}} \pubyear{2015}

\maketitle

\label{firstpage}

\begin{abstract}

V752 Cen is a triple-lined spectroscopic contact binary. Its multi-color light curves were obtained in the years 1971 and 2018, independently. Photometric analyses reveal that the two sets of light curves produce almost consistent results. It contains a W-subtype totally eclipsing binary, and its mass ratio and fill-out factor are $q = 3.35(1)$ and $f = 29(2)\,\%$. The absolute elements of its two component stars were determined to be $M_{1} = 0.39(2)M_\odot$, $M_{2} = 1.31(7)M_\odot$, $R_{1} = 0.77(1)R_\odot$, $R_{2} = 1.30(2)R_\odot$, $L_{1} = 0.75(3)L_\odot$ and $L_{2} = 2.00(7)L_\odot$. The period of V752 Cen is 0.37023198 day. The 0.37-d period remained constant from its first measurement in 1971 until the year 2000. However, it changed suddenly around the year 2000 and has been increasing continuously at a rate of $dP/dt=+5.05\times{10^{-7}}day\cdot year^{-1}$ since then, which can be explained by mass transfer from the less massive component star to the more massive one with a rate of $\frac{dM_{2}}{dt}=2.52\times{10^{-7}}M_\odot/year$. The period variation of V752 Cen over the 48 years in which the period has been monitored is really unusual, and is potentially related to effects from the possible presence of a nearby third star or of a pair of stars in a second binary.

\end{abstract}


\section{INTRODUCTION}

Eclipsing binaries have proved to be powerful probes for studying a wide range of astrophysical problems. They are the primary sources in providing fundamental stellar parameters such as mass, radius and luminosity, which are very important in testing theoretical models of stellar evolution \citep{2016AcA....66..405S}. Over the past decades, the number of known eclipsing binaries has increased rapidly. Photometric light curves of eclipsing binaries are archived by many survey projects, such as the Optical Gravitational Lensing Experiment (OGLE) \citep{2011AcA....61..103G,1997AJ....113.1112R}, the All Sky Automated Survey (ASAS) \citep{2019MNRAS.486.1907J} and the Kepler Space Telescope \citep{2011AJ....141...83P,2011AJ....142..160S}. What is even more exciting is that many spectroscopic survey projects are carried out at the same time, such as the Sloan Digital Sky Survey (SDSS) \citep{2000AJ....120.1579Y}, the Large Sky Area Multi-Object Fibre Spectroscopic Telescope (LAMOST) \citep{2017RAA....17...87Q,2018ApJS..235....5Q} and the Gaia mission \citep{2016A&A...595A...1G,2018A&A...616A...1G}. This means that many eclipsing binaries have been observed by both photometric and spectroscopic methods.

The observations on contact binaries have a long history. Almost all identified contact binaries belong to eclipsing binaries. However, the formation of contact binaries is still an open issue. Some researchers claim that contact binaries evolve from low-mass detached close binaries via the loss of angular momentum by means of a magnetic wind \citep{2004ARep...48..219T,2013MNRAS.430.2029Y}. Another hypothesis is that tertiary components have played a very important role in the formation of contact binaries by removing angular momentum from the central binaries \citep{2013ApJS..209...13Q,2014AJ....148...79Q}. Tertiary components are really common in contact binaries and almost every kind of celestial body can be a tertiary component orbiting the central binary, from planets to black holes \citep{2016PASJ...68..102X}. \citet{2006A&A...450..681T} conclude that $96\,\%$ of binaries with periods shorter than 3 days have tertiary components, based on a survey of 165 solar-type spectroscopic binaries. What is more, some contact binaries even have close-in companions with distance less than 3 AU \citep{2017PASJ...69...37Z}, and contact binaries with more than one companions were also reported \citep{2013AJ....145...39Z,2015AJ....149..120L}. All of these hierarchical contact binaries are very important samples for investigating the dynamic interactions among multi-stellar systems.

In 1970, \citet{1970PASP...82.1065B} reported three newly discovered W UMa type contact binaries, one of which was V752 Cen (HD 101799, $V$ = $9.1^{m}$). Later, \citet{1971IBVS..576....1S,1973AJ.....78..413S} observed V752 Cen photometrically and obtained $UBV$ light curves. They claimed that V752 Cen was a completely eclipsing binary system and its components nearly filled their respective Roche lobes (a semi-detached system). The radial velocity curves of V752 Cen were also published by \citet{1974AJ.....79..391S}, which revealed that V752 Cen was a double-lined eclipsing binary with the spectral type of its two component stars to be F8 and F5. However, \citet{1976PASP...88..936L} claimed that V752 Cen was a contact binary with a fill-out factor of $f = 6.3\,\%$ after he reanalyzed the light curves obtained by \citet{1971IBVS..576....1S,1973AJ.....78..413S}. \citet{1993ApJ...407..237B} also supported the contact configuration with a little higher fill-out factor ($9\,\%$). The fill-out factor is defined as $f = \frac{\Omega_{in}-\Omega_{1,2}}{\Omega_{in}-\Omega_{out}}$, where $\Omega_{in}$ and $\Omega_{out}$ are the potentials at the inner and outer Lagrangian points of a binary system, and $\Omega_{1,2}$ are the surface potentials of the component stars \citep{1973AcA....23...79R}. The high resolution spectroscopy of V752 Cen found that it may be a triple-lined spectroscopic quadruple system \citep{2009ASPC..404..199S}. \citet{2015JAVSO..43...38M} pointed out that the orbital period of V752 Cen was not always constant, and there were some small magnitude changes around the time that the period was inferred to change. V752 Cen was also listed in the Gaia Data Release 2 (DR2) with effective temperature to be $T_1 = 6138K$ and parallax to be $7.96 mas$ \citep{2016A&A...595A...1G,2018A&A...616A...1G}.

In the present work, we are going to reanalyze the light curves of V752 Cen published in April 1971 and the newly obtained ones in April 2018 to derive the physical parameters of the primary star and the secondary one. And also, its period variations over the past several decades will be revealed. Research on this hierarchical stellar system will provide valuable information about the evolution of hierarchical contact binary systems.

\section[]{PHOTOMETRIC OBSERVATIONS}

\begin{figure}
\begin{center}
\includegraphics[width=12cm]{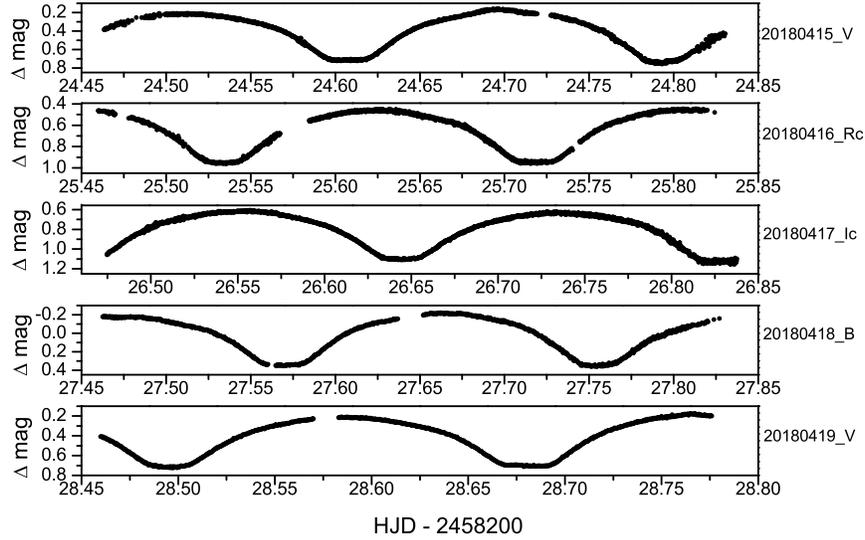}
\caption{Observational light curves of V752 Cen. The yellow, red, green and blue colors refer to $V$, $R_c$, $I_c$ and $B$ filters respectively. }\label{lc-obs}
\end{center}
\end{figure}

Photometric observations of V752 Cen were carried out at the Complejo Astronomico El Leoncito (CASLEO), San Juan, Argentina, with the 0.60 m Helen Sawyer Hogg (HSH) Telescope on 2018 April 15 - 19. Johnson-Cousins $BVR_cI_c$ filters were used during the observations. However, only one filter was used each night. The $V$ filter was used on 2018 April 15 and 19, and the $R_c$, $I_c$ and $B$ filters were used on 2018 April 16, 17 and 18, respectively. The observational light curves are displayed in Fig. \ref{lc-obs}. In order to get more minima times and cover a longer time span in the O - C diagram, V752 Cen was observed again with the $I_c$ filter on 2019 June 7. A total of ten mid-eclipse times of V752 Cen were determined, which are listed in Table \ref{New_minimum}.

\begin{table}\small
\begin{center}
\caption{Newly determined mid-eclipse times.}\label{New_minimum}
\begin{tabular}{cccccc}\hline\hline
    JD (Hel.)     &  Error (days)  &  p/s &    Filter   &   Method \\\hline\hline
  2458224.6073    & $\pm0.0001$    &   p  &      $V$    &    CCD   \\
  2458224.7936    & $\pm0.0001$    &   s  &      $V$    &    CCD   \\
  2458225.5337    & $\pm0.0001$    &   s  &     $R_C$   &    CCD   \\
  2458225.7181    & $\pm0.0001$    &   p  &     $R_C$   &    CCD   \\
  2458226.6443    & $\pm0.0001$    &   s  &     $I_C$   &    CCD   \\
  2458227.5692    & $\pm0.0001$    &   p  &      $B$    &    CCD   \\
  2458227.7551    & $\pm0.0001$    &   s  &      $B$    &    CCD   \\
  2458228.4956    & $\pm0.0001$    &   s  &      $V$    &    CCD   \\
  2458228.6803    & $\pm0.0001$    &   p  &      $V$    &    CCD   \\
  2458642.6058    & $\pm0.0001$    &   p  &      $I_C$  &    CCD   \\
\hline\hline
\end{tabular}
\end{center}
\end{table}

\section[]{INVESTIGATION OF PERIOD VARIATIONS}
Period variations commonly appear in close binary systems, especially for contact binaries, due to the possible mass transfer between their component stars or angular momentum loss from the binary systems. V752 Cen is a short period eclipsing binary system, which is listed in the International Variable Star Index (VSX)\footnote{https://www.aavso.org/vsx/} as having a period P = 0.37023198 days. In the present work, all published mid-eclipse times of V752 Cen were collected to investigate its period variations. And also, V752 Cen was observed by the SuperWASP project \citep{2010A&A...520L..10B} and  the Transiting Exoplanet Survey Satellite (TESS)\citep{2015JATIS...1a4003R}. The observational light curves were downloaded and 84 mid-eclipse times were determined from SuperWASP data. V752 Cen was observed by TESS from 2019 March 26 to 2019 April 22 continuously and 120 times of minima were determined (see Table 2). The minima in HJD were converted to BJD since the minima determined in TESS data used BJD time system. The linear equation used to calculate the $O - C$ values is:

\begin{equation}
Min.I(BJD) = 2456108.3448+0^{d}.37023198\times{E}.\label{Epoch_O-C}
\end{equation}

Column 1  - Heliocentric Julian Date of the observed mid-eclipse times (HJD - 2400000);

Column 2  - Barycentric Julian Date of the observed mid-eclipse times (BJD - 2400000);

Column 3  - primary (p) or secondary (s) mid-eclipse times;

Column 4  - PE, Vis and CCD refer to photoelectric, visual and Charge Coupled Device observations;

Column 5  - error of mid-eclipse times;

Column 6  - cycle numbers from the initial epoch;

Column 7  - the $O - C$ values calculated from Equation \ref{Epoch_O-C};

Column 8  - the references;

\begin{small}
\begin{longtable}{llllllll}
\caption{Mid-eclipse times and $O-C$ values for V752 Cen (the whole table is available in the online journal).}\label{OC}
\endfirsthead
\multicolumn{8}{l}{Table \ref{OC} continued }\\\hline\hline
     HJD       &        BJD       &  p/s     &   Method   &     Error    &    Epoch       &    $O - C$      & Ref.   \\
 (2400000+)    &    (2400000+)    &          &            &     (days)   &                &    (days)       &        \\\hline
\endhead
\hline\hline
\endfoot
\endlastfoot
\hline\hline
     HJD       &        BJD       &  p/s     &   Method   &     Error    &    Epoch       &    $O - C$      & Ref.   \\
 (2400000+)    &    (2400000+)    &          &            &     (days)   &                &    (days)       &        \\\hline
40648.613      &    40648.6135    &  s	      &     PE     &              &   -41757.5	   &    0.2306     &   1       \\
40648.98393    &    40648.9844    &  s	      &     PE     &              &   -41756.5	   &    0.2313     &   2       \\
41056.60633    &    41056.6068    &  s	      &     PE     &              &   -40655.5	   &    0.2283     &   2       \\
41056.60640    &    41056.6069    &  s	      &     PE     &              &   -40655.5	   &    0.2284     &   2       \\
41056.60652    &    41056.6070    &  s	      &     PE     &              &   -40655.5	   &    0.2285     &   2       \\
41057.53144    &    41057.5319    &  p	      &     PE     &              &   -40653  	   &    0.2278     &   2      \\
41057.53191    &    41057.5324    &  p	      &     PE     &              &   -40653  	   &    0.2283     &   2       \\
41057.53220    &    41057.5327    &  p	      &     PE     &              &   -40653  	   &    0.2286     &   2       \\
41057.71732    &    41057.7178    &  s	      &     PE     &              &   -40652.5	   &    0.2286     &   2       \\
41057.71742    &    41057.7179    &  s	      &     PE     &              &   -40652.5	   &    0.2287     &   2       \\
41057.71749    &    41057.7180    &  s	      &     PE     &              &   -40652.5	   &    0.2287     &   2       \\
41058.64242    &    41058.6429    &  p	      &     PE     &              &   -40650  	   &    0.2281     &   2       \\
41058.64279    &    41058.6433    &  p	      &     PE     &              &   -40650  	   &    0.2285     &   2       \\
41058.64353    &    41058.6440    &  p	      &     PE     &              &   -40650   	   &    0.2292     &   2       \\
41059.56740    &    41059.5679    &  s	      &     PE     &              &   -40647.5	   &    0.2275     &   2       \\
41059.56798    &    41059.5684    &  s	      &     PE     &              &   -40647.5	   &    0.2281     &   2       \\
41059.56809    &    41059.5686    &  s	      &     PE     &              &   -40647.5	   &    0.2282     &   2       \\
41059.75328    &    41059.7537    &  p	      &     PE     &              &   -40647  	   &    0.2283     &   2       \\
41059.75350    &    41059.7540    &  p	      &     PE     &              &   -40647  	   &    0.2285     &   2       \\
41059.75370    &    41059.7542    &  p	      &     PE     &              &   -40647  	   &    0.2287     &   2       \\
41060.49273    &    41060.4932    &  p	      &     PE     &              &   -40645  	   &    0.2272     &   2       \\
41060.49285    &    41060.4933    &  p	      &     PE     &              &   -40645  	   &    0.2274     &   2       \\
41060.49292    &    41060.4934    &  p	      &     PE     &              &   -40645  	   &    0.2274     &   2       \\
41060.67820    &    41060.6787    &  s	      &     PE     &              &   -40644.5	   &    0.2276     &   2       \\
41060.67834    &    41060.6788    &  s	      &     PE     &              &   -40644.5	   &    0.2277     &   2       \\
41060.67877    &    41060.6792    &  s	      &     PE     &              &   -40644.5	   &    0.2282     &   2       \\
41061.60298    &    41061.6034    &  p	      &     PE     &              &   -40642  	   &    0.2268     &   2       \\
41061.60383    &    41061.6043    &  p	      &     PE     &              &   -40642  	   &    0.2276     &   2       \\
41061.60395    &    41061.6044    &  p	      &     PE     &              &   -40642  	   &    0.2278     &   2       \\
41063.63958    &    41063.6400    &  s	      &     PE     &              &   -40636.5	   &    0.2271     &   2       \\
41063.63972    &    41063.6402    &  s	      &     PE     &              &   -40636.5	   &    0.2273     &   2       \\
41063.63973    &    41063.6402    &  s	      &     PE     &              &   -40636.5	   &    0.2273     &   2       \\
41064.56504    &    41064.5655    &  p	      &     PE     &              &   -40634  	   &    0.2270     &   2       \\
41064.56545    &    41064.5659    &  p	      &     PE     &              &   -40634  	   &    0.2274     &   2       \\
41064.56676    &    41064.5672    &  p	      &     PE     &              &   -40634  	   &    0.2287     &   2       \\
43244.457      &    43244.4576    &  p	      &     Vis    &              &   -34746  	   &    0.1932     &   3      \\
43245.378      &    43245.3786    &  s	      &     Vis    &              &   -34743.5	   &    0.1886     &   3      \\
43918.649      &    43918.6496    &  p	      &     Vis    &              &   -32925  	   &    0.1927     &   3      \\
43937.533      &    43937.5336    &  p	      &     Vis    &              &   -32874  	   &    0.1949     &   3      \\
44234.616      &    44234.6165    &  s	      &     Vis    &              &   -32071.5	   &    0.1667     &   3      \\
44235.547      &    44235.5476    &  p	      &     Vis    &              &   -32069  	   &    0.1721     &   3      \\
44243.681      &    44243.6816    &  p	      &     Vis    &              &   -32047   	   &    0.1610     &   3      \\
45428.406      &    45428.4066    &  p	      &     Vis    &              &   -28847  	   &    0.1437     &   3      \\
45430.456      &    45430.4566    &  s	      &     Vis    &              &   -28841.5	   &    0.1575     &   3      \\
45431.372      &    45431.3726    &  p	      &     Vis    &              &   -28839  	   &    0.1479     &   3      \\
45436.357      &    45436.3576    &  s	      &     Vis    &              &   -28825.5	   &    0.1347     &   3      \\
45438.400      &    45438.4006    &  p	      &     Vis    &              &   -28820  	   &    0.1415     &   3      \\
45753.466      &    45753.4666    &  p	      &     Vis    &              &   -27969  	   &    0.1401     &   3      \\
45804.379      &    45804.3796    &  s	      &     Vis    &              &   -27831.5	   &    0.1462     &   3      \\
45806.391      &    45806.3916    &  p	      &     Vis    &              &   -27826  	   &    0.1219     &   3      \\
46162.362      &    46162.3626    &  s	      &     Vis    &              &   -26864.5	   &    0.1149     &   3      \\
46163.305      &    46163.3056    &  p	      &     Vis    &              &   -26862  	   &    0.1323     &   3      \\
46164.417      &    46164.4176    &  p	      &     Vis    &              &   -26859  	   &    0.1336     &   3      \\
46166.277      &    46166.2776    &  p	      &     Vis    &              &   -26854  	   &    0.1424     &   3      \\
46167.383      &    46167.3836    &  p	      &     Vis    &              &   -26851  	   &    0.1377     &   3      \\
46530.385      &    46530.3857    &  s	      &     Vis    &              &   -25870.5	   &    0.1273     &   3      \\
46535.379      &    46535.3797    &  p	      &     Vis    &              &   -25857  	   &    0.1232     &   3      \\
46798.595      &    46798.5957    &  p	      &     Vis    &              &   -25146  	   &    0.1042     &   3      \\
46884.321      &    46884.3217    &  s	      &     Vis    &              &   -24914.5	   &    0.1215     &   3      \\
47262.325      &    47262.3257    &  s	      &     Vis    &              &   -23893.5	   &    0.1187     &   3      \\
47581.452      &    47581.4527    &  s	      &     Vis    &              &   -23031.5	   &    0.1057     &   3      \\
47663.272      &    47663.2727    &  s	      &     Vis    &              &   -22810.5	   &    0.1045     &   3      \\
48008.319      &    48008.3197    &  s	      &     Vis    &              &   -21878.5	   &    0.0953     &   3      \\
48373.360      &    48373.3607    &  s	      &     Vis    &     0.004    &   -20892.5	   &    0.0875     &   3      \\
49345.583      &    49345.5837    &  s	      &     Vis    &     0.005    &   -18266.5	   &    0.0814     &   3      \\
49773.920      &    49773.9207    &  s	      &     Vis    &     0.008    &   -17109.5	   &    0.0600     &   3      \\
52546.3492     &    52546.3500    &  p	      &     CCD    &              &   -9621  	     &    0.0070     &   4      \\
52546.5368     &    52546.5375    &  s	      &     CCD    &              &   -9620.5 	   &    0.0095     &   4      \\
53524.1272     &    53524.1279    &  p	      &     CCD    &              &   -6980  	     &    0.0024     &   4      \\
53524.3073     &    53524.3080    &  s	      &     CCD    &              &   -6979.5 	   &   -0.0027     &   4      \\
53860.4807     &    53860.4814    &  s       &     CCD    &     0.0002   &   -6071.5      &     0.0001     &   5      \\
53862.3329     &    53862.3327    &  s       &     CCD    &     0.0002   &   -6066.5      &     0.0002     &   5      \\
53863.2572     &    53863.2579    &  p       &     CCD    &     0.0002   &   -6064        &    -0.0002     &   5      \\
53863.4424     &    53863.4432    &  s       &     CCD    &     0.0002   &   -6063.5      &     0          &   5      \\
53864.3674     &    53864.3682    &  p       &     CCD    &     0.0002   &   -6061        &    -0.0006     &   5      \\
53870.2910     &    53870.2918    &  p       &     CCD    &     0.0002   &   -6045        &    -0.0007     &   5      \\
53880.2876     &    53880.2884    &  p       &     CCD    &     0.0002   &   -6018        &    -0.0004     &   5      \\
53881.3981     &    53881.3989    &  p       &     CCD    &     0.0002   &   -6015        &    -0.0006     &   5      \\
53886.3971     &    53886.3978    &  s       &     CCD    &     0.0002   &   -6001.5      &     0.0003     &   5      \\
53887.3216     &    53887.3223    &  p       &     CCD    &     0.0003   &   -5999        &    -0.0008     &   5      \\
53888.2482     &    53888.2489    &  s       &     CCD    &     0.0002   &   -5996.5      &     0.0002     &   5      \\
53890.2835     &    53890.2842    &  p       &     CCD    &     0.0002   &   -5991        &    -0.0008     &   5      \\
53892.3205     &    53892.3212    &  s       &     CCD    &     0.0002   &   -5985.5      &    -0.0001     &   5      \\
53896.3926     &    53896.3933    &  s       &     CCD    &     0.0001   &   -5974.5      &    -0.0005     &   5      \\
53897.3175     &    53897.3183    &  p       &     CCD    &     0.0001   &   -5972        &    -0.0011     &   5      \\
53898.2438     &    53898.2445    &  s       &     CCD    &     0.0002   &   -5969.5      &    -0.0005     &   5      \\
53902.3162     &    53902.3169    &  s       &     CCD    &     0.0001   &   -5958.5      &    -0.0006     &   5      \\
53904.3517     &    53904.3524    &  p       &     CCD    &     0.0002   &   -5953        &    -0.0014     &   5      \\
53905.2782     &    53905.2789    &  s       &     CCD    &     0.0002   &   -5950.5      &    -0.0005     &   5      \\
53907.3135     &    53907.3142    &  p       &     CCD    &     0.0002   &   -5945        &    -0.0015     &   5      \\
53910.2755     &    53910.2762    &  p       &     CCD    &     0.0002   &   -5937        &    -0.0013     &   5      \\
53917.3099     &    53917.3107    &  p       &     CCD    &     0.0003   &   -5918        &    -0.0013     &   5      \\
54132.5998     &    54132.6005    &  s       &     CCD    &     0.0001   &   -5336.5      &    -0.0013     &   5      \\
54136.4867     &    54136.4875    &  p       &     CCD    &     0.0002   &   -5326        &    -0.0018     &   5      \\
54140.5594     &    54140.5601    &  p       &     CCD    &     0.0003   &   -5315        &    -0.0017     &   5      \\
54145.5585     &    54145.5592    &  s       &     CCD    &     0.0001   &   -5301.5      &    -0.0007     &   5      \\
54148.5203     &    54148.5211    &  s       &     CCD    &     0.0002   &   -5293.5      &    -0.0007     &   5      \\
54151.4817     &    54151.4825    &  s       &     CCD    &     0.0002   &   -5285.5      &    -0.0012     &   5      \\
54153.5172     &    54153.5180    &  p       &     CCD    &     0.0002   &   -5280        &    -0.0020     &   5      \\
54155.5552     &    54155.5559    &  s       &     CCD    &     0.0002   &   -5274.5      &    -0.0003     &   5      \\
54159.4407     &    54159.4414    &  p       &     CCD    &     0.0002   &   -5264        &    -0.0023     &   5      \\
54168.5139     &    54168.5146    &  s       &     CCD    &     0.0001   &   -5239.5      &     0.0003     &   5      \\
54169.4369     &    54169.4376    &  p       &     CCD    &     0.0003   &   -5237        &    -0.0023     &   5      \\
54169.6248     &    54169.6256    &  s       &     CCD    &     0.0002   &   -5236.5      &     0.0005     &   5      \\
54170.3648     &    54170.3655    &  s       &     CCD    &     0.0003   &   -5234.5      &     0          &   5      \\
54170.5480     &    54170.5487    &  p       &     CCD    &     0.0003   &   -5234        &    -0.0019     &   5      \\
54175.3611     &    54175.3619    &  p       &     CCD    &     0.0003   &   -5221        &    -0.0018     &   5      \\
54176.4718     &    54176.4725    &  p       &     CCD    &     0.0002   &   -5218        &    -0.0018     &   5      \\
54178.3229     &    54178.3236    &  p       &     CCD    &     0.0003   &   -5213        &    -0.0018     &   5      \\
54179.6203     &    54179.6210    &  s       &     CCD    &     0.0002   &   -5209.5      &    -0.0003     &   5      \\
54180.3606     &    54180.3613    &  s       &     CCD    &     0.0002   &   -5207.5      &    -0.0004     &   5      \\
54180.5442     &    54180.5449    &  p       &     CCD    &     0.0002   &   -5207        &    -0.0019     &   5      \\
54181.4711     &    54181.4718    &  s       &     CCD    &     0.0001   &   -5204.5      &    -0.0006     &   5      \\
54183.3222     &    54183.3229    &  s       &     CCD    &     0.0001   &   -5199.5      &    -0.0007     &   5      \\
54184.4333     &    54184.4340    &  s       &     CCD    &     0.0002   &   -5196.5      &    -0.0003     &   5      \\
54199.4263     &    54199.4271    &  p       &     CCD    &     0.0002   &   -5156        &    -0.0016     &   5      \\
54200.3521     &    54200.3528    &  s       &     CCD    &     0.0002   &   -5153.5      &    -0.0015     &   5      \\
54200.5367     &    54200.5375    &  p       &     CCD    &     0.0002   &   -5153        &    -0.0019     &   5      \\
54201.2776     &    54201.2783    &  p       &     CCD    &     0.0002   &   -5151        &    -0.0016     &   5      \\
54203.3139     &    54203.3146    &  s       &     CCD    &     0.0001   &   -5145.5      &    -0.0015     &   5      \\
54205.3500     &    54205.3507    &  p       &     CCD    &     0.0002   &   -5140        &    -0.0017     &   5      \\
54206.2758     &    54206.2765    &  s       &     CCD    &     0.0002   &   -5137.5      &    -0.0014     &   5      \\
54207.3864     &    54207.3872    &  s       &     CCD    &     0.0002   &   -5134.5      &    -0.0015     &   5      \\
54208.3115     &    54208.3122    &  p       &     CCD    &     0.0002   &   -5132        &    -0.0021     &   5      \\
54211.2732     &    54211.2740    &  p       &     CCD    &     0.0002   &   -5124        &    -0.0022     &   5      \\
54211.4592     &    54211.4599    &  s       &     CCD    &     0.0002   &   -5123.5      &    -0.0013     &   5      \\
54214.4216     &    54214.4223    &  s       &     CCD    &     0.0002   &   -5115.5      &    -0.0008     &   5      \\
54216.2724     &    54216.2731    &  s       &     CCD    &     0.0001   &   -5110.5      &    -0.0011     &   5      \\
54225.3426     &    54225.3433    &  p       &     CCD    &     0.0002   &   -5086        &    -0.0016     &   5      \\
54231.2660     &    54231.2667    &  p       &     CCD    &     0.0002   &   -5070        &    -0.0019     &   5      \\
54232.3766     &    54232.3773    &  p       &     CCD    &     0.0002   &   -5067        &    -0.0020     &   5      \\
54233.3038     &    54233.3045    &  s       &     CCD    &     0.0001   &   -5064.5      &    -0.0004     &   5      \\
54238.3003     &    54238.3011    &  p       &     CCD    &     0.0002   &   -5051        &    -0.0020     &   5      \\
54246.2616     &    54246.2623    &  s       &     CCD    &     0.0002   &   -5029.5      &    -0.0007     &   5      \\
54254.2189     &    54254.2196    &  p       &     CCD    &     0.0003   &   -5008        &    -0.0034     &   5      \\
54254.4052     &    54254.4060    &  s       &     CCD    &     0.0002   &   -5007.5      &    -0.0022     &   5      \\
54265.3273     &    54265.3281    &  p       &     CCD    &     0.0002   &   -4978        &    -0.0019     &   5      \\
54266.2535     &    54266.2543    &  s       &     CCD    &     0.0001   &   -4975.5      &    -0.0013     &   5      \\
54268.2896     &    54268.2904    &  p       &     CCD    &     0.0002   &   -4970        &    -0.0015     &   5      \\
54271.2512     &    54271.2520    &  p       &     CCD    &     0.0002   &   -4962        &    -0.0017     &   5      \\
54273.2883     &    54273.2890    &  s       &     CCD    &     0.0001   &   -4956.5      &    -0.0010     &   5      \\
54484.5038     &    54484.5045    &  p       &     CCD    &     0.0002   &   -4386        &    -0.0028     &   5      \\
54497.4625     &    54497.4632    &  p       &     CCD    &     0.0002   &   -4351        &    -0.0022     &   5      \\
54499.4993     &    54499.5000    &  s       &     CCD    &     0.0001   &   -4345.5      &    -0.0017     &   5      \\
54502.4609     &    54502.4616    &  s       &     CCD    &     0.0002   &   -4337.5      &    -0.0020     &   5      \\
54525.4154     &    54525.4161    &  s       &     CCD    &     0.0001   &   -4275.5      &    -0.0018     &   5      \\
54531.1528     &    54531.1536    &  p	     &     CCD    &              &   -4260  	    &    -0.0030     &   4      \\
54531.3398     &    54531.3406    &  s	     &     CCD    &              &   -4259.5 	    &    -0.0011     &   4      \\
54536.3364     &    54536.3372    &  p       &     CCD    &     0.0002   &   -4246        &    -0.0026     &   5      \\
54564.2897     &    54564.2904    &  s       &     CCD    &     0.0001   &   -4170.5      &    -0.0019     &   5      \\
54566.3250     &    54566.3257    &  p       &     CCD    &     0.0002   &   -4165        &    -0.0029     &   5      \\
54569.2869     &    54569.2877    &  p       &     CCD    &     0.0002   &   -4157        &    -0.0028     &   5      \\
54571.3233     &    54571.3241    &  s       &     CCD    &     0.0002   &   -4151.5      &    -0.0026     &   5      \\
54579.2827     &    54579.2835    &  p       &     CCD    &     0.0002   &   -4130        &    -0.0032     &   5      \\
54581.5041     &    54581.5048    &  p       &     CCD    &     0.0003   &   -4124        &    -0.0033     &   5      \\
54592.2414     &    54592.2422    &  p       &     CCD    &     0.0002   &   -4095        &    -0.0026     &   5      \\
54988.390      &    54988.3908    &  p	     &     CCD    &     0.006    &   -3025  	    &    -0.0023     &   6      \\
55395.277      &    55395.2778    &  p	     &     CCD    &     0.005    &   -1926  	    &    -0.0002     &   7      \\
56108.3440     &    56108.3448    &  p	     &     CCD    &     0.0030   &   0            &    0           &   8      \\
57051.1458     &    57051.1466    &  s	     &     CCD    &     0.0002   &   2546.5 	    &    0.0061      &   4      \\
58224.6073     &    58224.6081    &  p	     &     CCD    &     0.0001   &   5716   	    &    0.0173      &   9      \\
58224.7936     &    58224.7944    &  s	     &     CCD    &     0.0001   &   5716.5  	    &    0.0185      &   9      \\
58225.5337     &    58225.5345    &  s	     &     CCD    &     0.0001   &   5718.5 	    &    0.0181      &   9      \\
58225.7181     &    58225.7189    &  p	     &     CCD    &     0.0001   &   5719   	    &    0.0174      &   9      \\
58226.6443     &    58226.6451    &  s	     &     CCD    &     0.0001   &   5721.5 	    &    0.0180      &   9      \\
58227.5692     &    58227.5700    &  p	     &     CCD    &     0.0001   &   5724   	    &    0.0174      &   9      \\
58227.7551     &    58227.7559    &  s	     &     CCD    &     0.0001   &   5724.5 	    &    0.0181      &   9      \\
58228.4956     &    58228.4964    &  s	     &     CCD    &     0.0001   &   5726.5 	    &    0.0182      &   9      \\
58228.6803     &    58228.6811    &  p	     &     CCD    &     0.0001   &   5727   	    &    0.0178      &   9      \\
               &    58570.7804    &  p       &     CCD    &     0.0001   &   6651         &    0.0227      &   10     \\
               &    58570.9653    &  s       &     CCD    &     0.0001   &   6651.5       &    0.0225      &   10     \\
               &    58571.1507    &  p       &     CCD    &     0.0001   &   6652         &    0.0227      &   10     \\
               &    58571.3356    &  s       &     CCD    &     0.0001   &   6652.5       &    0.0225      &   10     \\
               &    58571.5209    &  p       &     CCD    &     0.0001   &   6653         &    0.0227      &   10     \\
               &    58571.7058    &  s       &     CCD    &     0.0001   &   6653.5       &    0.0225      &   10     \\
               &    58571.8911    &  p       &     CCD    &     0.0001   &   6654         &    0.0227      &   10     \\
               &    58572.0761    &  s       &     CCD    &     0.0001   &   6654.5       &    0.0226      &   10     \\
               &    58572.2614    &  p       &     CCD    &     0.0001   &   6655         &    0.0228      &   10     \\
               &    58572.4463    &  s       &     CCD    &     0.0001   &   6655.5       &    0.0226      &   10     \\
               &    58572.8165    &  s       &     CCD    &     0.0001   &   6656.5       &    0.0226      &   10     \\
               &    58573.0018    &  p       &     CCD    &     0.0001   &   6657         &    0.0228      &   10     \\
               &    58573.1867    &  s       &     CCD    &     0.0001   &   6657.5       &    0.0225      &   10     \\
               &    58573.3720    &  p       &     CCD    &     0.0001   &   6658         &    0.0227      &   10     \\
               &    58573.5570    &  s       &     CCD    &     0.0001   &   6658.5       &    0.0226      &   10     \\
               &    58573.7422    &  p       &     CCD    &     0.0001   &   6659         &    0.0227      &   10     \\
               &    58573.9273    &  s       &     CCD    &     0.0001   &   6659.5       &    0.0226      &   10     \\
               &    58574.1125    &  p       &     CCD    &     0.0001   &   6660         &    0.0227      &   10     \\
               &    58574.2975    &  s       &     CCD    &     0.0001   &   6660.5       &    0.0227      &   10     \\
               &    58574.4827    &  p       &     CCD    &     0.0001   &   6661         &    0.0227      &   10     \\
               &    58574.6678    &  s       &     CCD    &     0.0001   &   6661.5       &    0.0227      &   10     \\
               &    58574.8529    &  p       &     CCD    &     0.0001   &   6662         &    0.0227      &   10     \\
               &    58575.0380    &  s       &     CCD    &     0.0001   &   6662.5       &    0.0227      &   10     \\
               &    58575.2231    &  p       &     CCD    &     0.0001   &   6663         &    0.0226      &   10     \\
               &    58575.4083    &  s       &     CCD    &     0.0001   &   6663.5       &    0.0227      &   10     \\
               &    58575.5933    &  p       &     CCD    &     0.0001   &   6664         &    0.0226      &   10     \\
               &    58575.7785    &  s       &     CCD    &     0.0001   &   6664.5       &    0.0227      &   10     \\
               &    58575.9636    &  p       &     CCD    &     0.0001   &   6665         &    0.0226      &   10     \\
               &    58576.1487    &  s       &     CCD    &     0.0001   &   6665.5       &    0.0227      &   10     \\
               &    58576.3338    &  p       &     CCD    &     0.0001   &   6666         &    0.0226      &   10     \\
               &    58576.5189    &  s       &     CCD    &     0.0001   &   6666.5       &    0.0226      &   10     \\
               &    58576.7040    &  p       &     CCD    &     0.0001   &   6667         &    0.0226      &   10     \\
               &    58576.8891    &  s       &     CCD    &     0.0001   &   6667.5       &    0.0220      &   10     \\
               &    58577.0743    &  p       &     CCD    &     0.0001   &   6668         &    0.0226      &   10     \\
               &    58577.2594    &  s       &     CCD    &     0.0001   &   6668.5       &    0.0226      &   10     \\
               &    58577.4445    &  p       &     CCD    &     0.0001   &   6669         &    0.0227      &   10     \\
               &    58577.6296    &  s       &     CCD    &     0.0001   &   6669.5       &    0.0226      &   10     \\
               &    58577.8147    &  p       &     CCD    &     0.0001   &   6670         &    0.0226      &   10     \\
               &    58577.9998    &  s       &     CCD    &     0.0001   &   6670.5       &    0.0226      &   10     \\
               &    58578.1850    &  p       &     CCD    &     0.0001   &   6671         &    0.0227      &   10     \\
               &    58578.3701    &  s       &     CCD    &     0.0001   &   6671.5       &    0.0226      &   10     \\
               &    58578.5552    &  p       &     CCD    &     0.0001   &   6672         &    0.0226      &   10     \\
               &    58578.7403    &  s       &     CCD    &     0.0001   &   6672.5       &    0.0226      &   10     \\
               &    58578.9255    &  p       &     CCD    &     0.0001   &   6673         &    0.0227      &   10     \\
               &    58579.1105    &  s       &     CCD    &     0.0001   &   6673.5       &    0.0226      &   10     \\
               &    58579.2956    &  p       &     CCD    &     0.0001   &   6674         &    0.0226      &   10     \\
               &    58579.4807    &  s       &     CCD    &     0.0001   &   6674.5       &    0.0226      &   10     \\
               &    58579.6659    &  p       &     CCD    &     0.0001   &   6675         &    0.0227      &   10     \\
               &    58579.8509    &  s       &     CCD    &     0.0001   &   6675.5       &    0.0226      &   10     \\
               &    58580.0361    &  p       &     CCD    &     0.0001   &   6676         &    0.0227      &   10     \\
               &    58580.2212    &  s       &     CCD    &     0.0001   &   6676.5       &    0.0226      &   10     \\
               &    58580.4064    &  p       &     CCD    &     0.0001   &   6677         &    0.0227      &   10     \\
               &    58580.5915    &  s       &     CCD    &     0.0001   &   6677.5       &    0.0226      &   10     \\
               &    58580.7767    &  p       &     CCD    &     0.0001   &   6678         &    0.0227      &   10     \\
               &    58580.9617    &  s       &     CCD    &     0.0001   &   6678.5       &    0.0227      &   10     \\
               &    58581.1469    &  p       &     CCD    &     0.0001   &   6679         &    0.0227      &   10     \\
               &    58581.3320    &  s       &     CCD    &     0.0001   &   6679.5       &    0.0227      &   10     \\
               &    58581.5171    &  p       &     CCD    &     0.0001   &   6680         &    0.0227      &   10     \\
               &    58581.7022    &  s       &     CCD    &     0.0001   &   6680.5       &    0.0227      &   10     \\
               &    58584.4793    &  p       &     CCD    &     0.0001   &   6688         &    0.0230      &   10     \\
               &    58584.6641    &  s       &     CCD    &     0.0001   &   6688.5       &    0.0228      &   10     \\
               &    58584.8494    &  p       &     CCD    &     0.0001   &   6689         &    0.0229      &   10     \\
               &    58585.0343    &  s       &     CCD    &     0.0001   &   6689.5       &    0.0227      &   10     \\
               &    58585.2197    &  p       &     CCD    &     0.0001   &   6690         &    0.0230      &   10     \\
               &    58585.4046    &  s       &     CCD    &     0.0001   &   6690.5       &    0.0227      &   10     \\
               &    58585.5900    &  p       &     CCD    &     0.0001   &   6691         &    0.0230      &   10     \\
               &    58585.7748    &  s       &     CCD    &     0.0001   &   6691.5       &    0.0227      &   10     \\
               &    58585.9602    &  p       &     CCD    &     0.0001   &   6692         &    0.0230      &   10     \\
               &    58586.1446    &  s       &     CCD    &     0.0001   &   6692.5       &    0.0223      &   10     \\
               &    58586.3304    &  p       &     CCD    &     0.0001   &   6693         &    0.0230      &   10     \\
               &    58586.5152    &  s       &     CCD    &     0.0001   &   6693.5       &    0.0227      &   10     \\
               &    58586.7007    &  p       &     CCD    &     0.0001   &   6694         &    0.0230      &   10     \\
               &    58586.8855    &  s       &     CCD    &     0.0001   &   6694.5       &    0.0227      &   10     \\
               &    58587.0709    &  p       &     CCD    &     0.0001   &   6695         &    0.0230      &   10     \\
               &    58587.2558    &  s       &     CCD    &     0.0001   &   6695.5       &    0.0228      &   10     \\
               &    58587.4411    &  p       &     CCD    &     0.0001   &   6696         &    0.0230      &   10     \\
               &    58587.6260    &  s       &     CCD    &     0.0001   &   6696.5       &    0.0228      &   10     \\
               &    58587.8114    &  p       &     CCD    &     0.0001   &   6697         &    0.0230      &   10     \\
               &    58587.9962    &  s       &     CCD    &     0.0001   &   6697.5       &    0.0227      &   10     \\
               &    58588.1816    &  p       &     CCD    &     0.0001   &   6698         &    0.0230      &   10     \\
               &    58588.3664    &  s       &     CCD    &     0.0001   &   6698.5       &    0.0227      &   10     \\
               &    58588.5519    &  p       &     CCD    &     0.0001   &   6699         &    0.0230      &   10     \\
               &    58588.7367    &  s       &     CCD    &     0.0001   &   6699.5       &    0.0227      &   10     \\
               &    58588.9221    &  p       &     CCD    &     0.0001   &   6700         &    0.0231      &   10     \\
               &    58589.1069    &  s       &     CCD    &     0.0001   &   6700.5       &    0.0228      &   10     \\
               &    58589.2923    &  p       &     CCD    &     0.0001   &   6701         &    0.0230      &   10     \\
               &    58589.4768    &  s       &     CCD    &     0.0001   &   6701.5       &    0.0224      &   10     \\
               &    58589.6625    &  p       &     CCD    &     0.0001   &   6702         &    0.0230      &   10     \\
               &    58589.8474    &  s       &     CCD    &     0.0001   &   6702.5       &    0.0228      &   10     \\
               &    58590.0328    &  p       &     CCD    &     0.0001   &   6703         &    0.0230      &   10     \\
               &    58590.2176    &  s       &     CCD    &     0.0001   &   6703.5       &    0.0228      &   10     \\
               &    58590.4031    &  p       &     CCD    &     0.0001   &   6704         &    0.0231      &   10     \\
               &    58590.5878    &  s       &     CCD    &     0.0001   &   6704.5       &    0.0227      &   10     \\
               &    58590.7733    &  p       &     CCD    &     0.0001   &   6705         &    0.0231      &   10     \\
               &    58590.9580    &  s       &     CCD    &     0.0001   &   6705.5       &    0.0227      &   10     \\
               &    58591.1435    &  p       &     CCD    &     0.0001   &   6706         &    0.0231      &   10     \\
               &    58591.3283    &  s       &     CCD    &     0.0001   &   6706.5       &    0.0227      &   10     \\
               &    58591.5138    &  p       &     CCD    &     0.0001   &   6707         &    0.0231      &   10     \\
               &    58591.6985    &  s       &     CCD    &     0.0001   &   6707.5       &    0.0227      &   10     \\
               &    58591.8840    &  p       &     CCD    &     0.0001   &   6708         &    0.0231      &   10     \\
               &    58592.0687    &  s       &     CCD    &     0.0001   &   6708.5       &    0.0227      &   10     \\
               &    58592.2543    &  p       &     CCD    &     0.0001   &   6709         &    0.0232      &   10     \\
               &    58592.4389    &  s       &     CCD    &     0.0001   &   6709.5       &    0.0227      &   10     \\
               &    58592.6245    &  p       &     CCD    &     0.0001   &   6710         &    0.0232      &   10     \\
               &    58592.8092    &  s       &     CCD    &     0.0001   &   6710.5       &    0.0227      &   10     \\
               &    58592.9948    &  p       &     CCD    &     0.0001   &   6711         &    0.0232      &   10     \\
               &    58593.1794    &  s       &     CCD    &     0.0001   &   6711.5       &    0.0227      &   10     \\
               &    58593.3650    &  p       &     CCD    &     0.0001   &   6712         &    0.0232      &   10     \\
               &    58593.5497    &  s       &     CCD    &     0.0001   &   6712.5       &    0.0228      &   10     \\
               &    58593.7352    &  p       &     CCD    &     0.0001   &   6713         &    0.0232      &   10     \\
               &    58593.9199    &  s       &     CCD    &     0.0001   &   6713.5       &    0.0228      &   10     \\
               &    58594.1055    &  p       &     CCD    &     0.0001   &   6714         &    0.0232      &   10     \\
               &    58594.2902    &  s       &     CCD    &     0.0001   &   6714.5       &    0.0228      &   10     \\
               &    58594.4757    &  p       &     CCD    &     0.0001   &   6715         &    0.0231      &   10     \\
               &    58594.6605    &  s       &     CCD    &     0.0001   &   6715.5       &    0.0228      &   10     \\
               &    58594.8460    &  p       &     CCD    &     0.0001   &   6716         &    0.0232      &   10     \\
               &    58595.0307    &  s       &     CCD    &     0.0001   &   6716.5       &    0.0228      &   10     \\
               &    58595.2162    &  p       &     CCD    &     0.0001   &   6717         &    0.0232      &   10     \\
               &    58595.4009    &  s       &     CCD    &     0.0001   &   6717.5       &    0.0228      &   10     \\
               &    58595.5864    &  p       &     CCD    &     0.0001   &   6718         &    0.0231      &   10     \\
58642.6058     &    58642.6066    &  p	     &     CCD    &     0.0001   &   6845   	    &    0.0239      &   9      \\\hline
\end{longtable}
\textbf
{\footnotesize References:} \footnotesize (1) \citet{1970PASP...82.1065B}; (2) \citet{1973AJ.....78..413S}; (3) BBSAG Bulletins\footnote{https://www.astronomie.info/bbsag/bulletins.html}; (4) \citet{2015JAVSO..43...38M}; (5) SuperWASP\footnote{https://wasp.cerit-sc.cz/form};  (6) \citet{2009OEJV..116....1P}; (7) \citet{2010OEJV..130....1P}; (8) \citet{2013OEJV..155....1P}; (9) The present work; (10) TESS.
\end{small}

\begin{figure}
\begin{center}
\includegraphics[width=12cm]{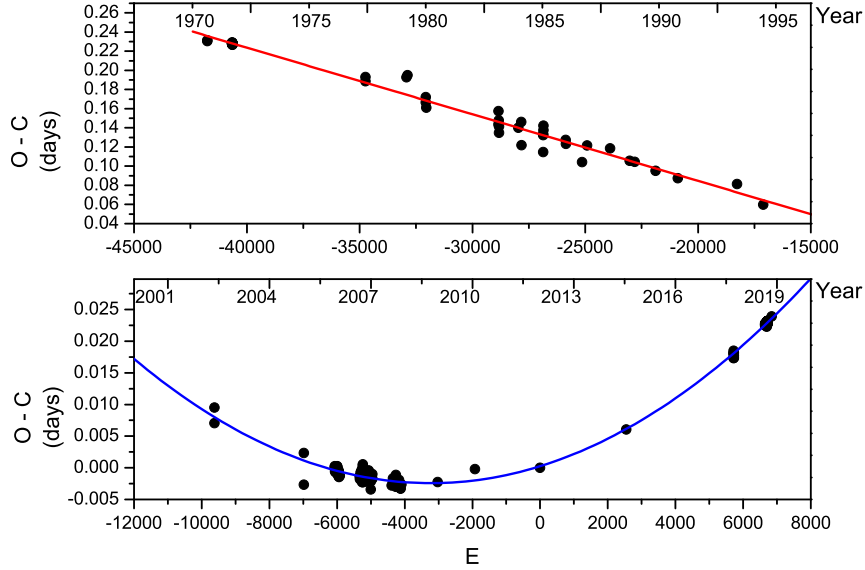}
\end{center}
\caption{In both panels, the solid circles are O-C values calculated from Equation \ref{Epoch_O-C}.  In the upper panel, the red line represents, relative to Equation \ref{Epoch_O-C}, the expected O-C curve based on Equation \ref{New_Epoch1}.  In the lower panel, the blue curve represents,  relative to Equation \ref{Epoch_O-C}, the expected O-C curve based on Equation \ref{New_Epoch2}.}\label{OC_fitting}
\end{figure}

The $O - C$ values calculated with Equation \ref{Epoch_O-C} are displayed in Fig. \ref{OC_fitting} with solid circles. The solid circles are split into two parts since the period of V752 Cen seems to change suddenly around the year 2000. Its period was constant during the first 30 years after it was first reported by \citet{1970PASP...82.1065B}, as displayed in the upper panel in Fig. \ref{OC_fitting}. The new linear equation is:
\begin{equation}
\begin{array}{lll}
Min.I(BJD) = 2456108.2896(\pm0.0002)+0^{d}.370225027(\pm0.000000005)\times{E}.\label{New_Epoch1}
\end{array}
\end{equation}

The second part of the $O - C$ values as shown in the lower part of Fig. \ref{OC_fitting}, is increasing continuously from the year 2000 up to the present day. Thus, a quadratic term is superposed on the initial linear Equation \ref{Epoch_O-C}. The new ephemeris is:

\begin{equation}
\begin{array}{lll}
Min.I(BJD) = 2456108.3443(\pm0.0001)+0^{d}.370233636(\pm0.000000004)\times{E}
         \\+2.56(\pm0.02)\times{10^{-10}}\times{E^{2}}. \label{New_Epoch2}
\end{array}
\end{equation}

\begin{figure}
\begin{center}
\includegraphics[width=12cm]{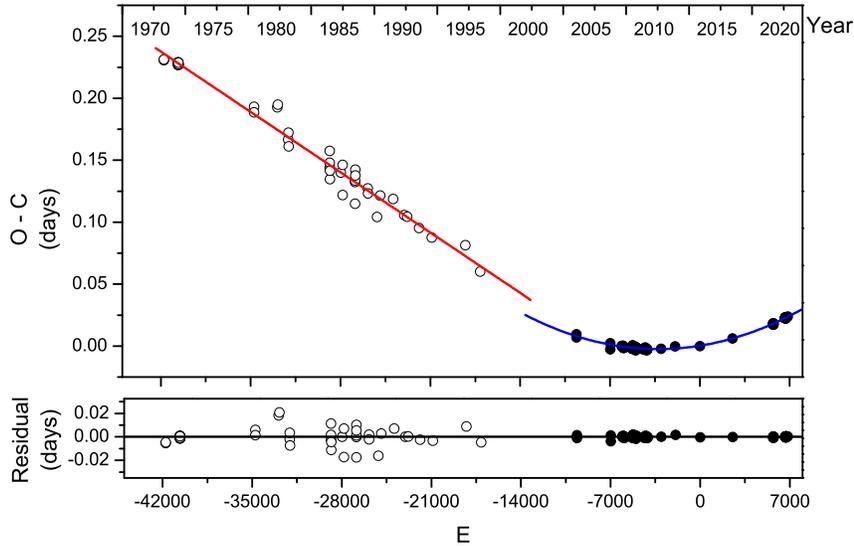}
\end{center}
\caption{The O - C curve for the whole 48 - year interval.}\label{OC_fitting-com}
\end{figure}

The fitting of the observed O - C data reveals that the period of V752 Cen changed suddenly around the year 2000. Since then, it has been increasing continuously at a rate of $dP/dt=+5.05\times{10^{-7}}day\cdot year^{-1}$ ($+0.044 s\cdot year^{-1}$). To illustrate the sudden period change more clearly, the O - C curve for the whole time span is shown in Fig. \ref{OC_fitting-com}. It should be mentioned that we have tried to fit the observed O - C data with other possibilities, for example, orbital motion of the contact binary in a long-period elliptical orbit. However, the latter possibility would require the presence of a star (or stars) with an implausibly high mass (masses), so it was dropped from further consideration.

\section{MODELING THE LIGHT CURVES}

\begin{figure}
\begin{center}
\includegraphics[width=12cm]{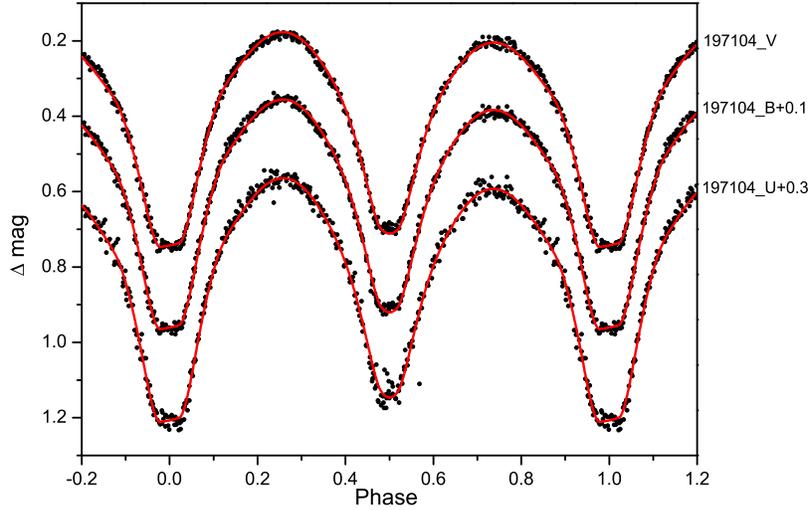}
\end{center}
\caption{The solid circles are the $U$, $B$ and $V$ band light curves observed in April 1971 and the red lines are the corresponding theoretical light curves.}\label{LC-1971}
\end{figure}

\begin{figure}
\begin{center}
\includegraphics[width=12cm]{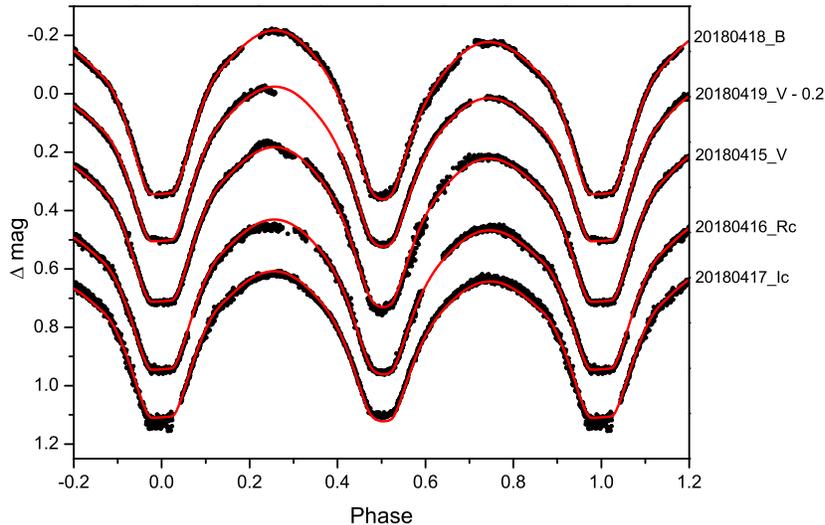}
\end{center}
\caption{The solid circles are the $B$, $V$, $R_c$and $I_c$ band light curves observed in April 2018 and the red lines are the corresponding theoretical light curves.}\label{LC-2018}
\end{figure}

The light curves were obtained in April, 1971. The newly observed light curves displayed in Fig. \ref{lc-obs} were obtained in April, 2018. The two sets of light curves are modeled with the Wilson-Devinney (W-D) program \citep{2012AJ....144...73W,2014ApJ...780..151W}. The light curves show that V752 Cen is a totally eclipsed W UMa system. Mode 3 for contact binary is chosen, and the q-search method is not necessary \citep{2005Ap&SS.296..221T}. The mean surface temperature is given as $T_1 = 6138K$ in the Gaia DR2 with a typical error of 324K \citep{2018A&A...616A...8A}. \citet{1974AJ.....79..391S} reported that the spectral type of the primary star is F8, corresponding to a temperature of 6250K \citep{Cox2000}. The spectral types derived from B - V and J - H were G0 and F7, respectively. We adopted the temperature from the Gaia DR2 since the uncertainty of about 300K in temperature will not significantly change parameters like mass ratio, orbital inclination and fill-out factor \citep{2015AJ....149..169J,2015AJ....150...83Z}. The gravity-darkening coefficients and bolometric albedo coefficients are set at $g_{1} = g_{2} = 0.32$ \citep{1967ZA.....65...89L} and $A_{1} = A_{2} = 0.5$ \citep{1969AcA....19..245R}, respectively, and the limb darkening coefficients are set accordingly \citep{1993AJ....106.2096V}. The star eclipsed at primary minimum is assumed to be the primary star. The adjustable parameters are the mass ratio ($q$), orbital inclination ($i$), modified dimensionless surface potential of the primary star ($\Omega_{1}$), mean surface temperature of the secondary star ($T_{2}$), bandpass luminosities of the primary star ($L_{1}$), and spots' parameters. The determined photometric parameters are listed in Table \ref{WD_results}. The theoretical light curves based on the Roche model are displayed in Fig. \ref{LC-1971} and Fig. \ref{LC-2018}. The $U$ and $B$ band light curves are shifted vertically by 0.3 and 0.1 mag in Fig. \ref{LC-1971}. The $V$ band light curves observed on 19 April 2018 are shifted vertically by -0.2 mag in Fig. \ref{LC-2018}.

\begin{table}
\begin{center}
\caption{Photometric solutions for V752 Cen}\label{WD_results}
\small
\begin{tabular}{lllllllll}
\hline\hline
Parameters                            &   Data 1971                     &    Data 2018        \\\hline
$T_{1}(K)   $                         &  6138(fixed)                     &  6138(fixed)      \\
q ($M_2/M_1$ )                        &  3.31($\pm0.02$)                 &  3.35($\pm0.01$)         \\
$i(^{\circ})$                         &  81.8($\pm0.2$)                  &  82.07($\pm0.06$)      \\
$\Omega_{in}$                         &  7.02                            &  7.08      \\
$\Omega_{out}$                        &  6.40                            &  6.45     \\
$\Omega_{1}=\Omega_{2}$               &  6.85($\pm0.02$)                 &  6.89($\pm0.01$)    \\
$T_{2}(K)$                            &  5947($\pm7$)                    &  6014($\pm2$)       \\
$\Delta T(K)$                         &  191                             &  124       \\
$T_{2}/T_{1}$                         &  0.969($\pm0.001$)               &  0.9798($\pm0.0003$)     \\
$L_{1}/(L_{1}+L_{2}$) ($U$)           &  0.305($\pm0.001$)               &                      \\
$L_{1}/(L_{1}+L_{2}$) ($B$)           &  0.296($\pm0.001$)               &  0.2798($\pm0.0003$)     \\
$L_{1}/(L_{1}+L_{2}$) ($V$)           &  0.286($\pm0.001$)               &  0.2735($\pm0.0003$)     \\
$L_{1}/(L_{1}+L_{2}$) ($V$)           &                                  &  0.2735($\pm0.0002$)     \\
$L_{1}/(L_{1}+L_{2}$) ($R_c$)         &                                  &  0.2706($\pm0.0003$)     \\
$L_{1}/(L_{1}+L_{2}$) ($I_c$)         &                                  &  0.2683($\pm0.0003$)     \\
$r_{1}(pole)$                         &  0.2733($\pm0.0005$)             &  0.2730($\pm0.0003$)     \\
$r_{1}(side)$                         &  0.2862($\pm0.0006$)             &  0.2861($\pm0.0003$)     \\
$r_{1}(back)$                         &  0.3289($\pm0.0011$)             &  0.3295($\pm0.0004$)     \\
$r_{2}(pole)$                         &  0.4664($\pm0.0020$)             &  0.4676($\pm0.0008$)     \\
$r_{2}(side)$                         &  0.5045($\pm0.0028$)             &  0.5059($\pm0.0012$)     \\
$r_{2}(back)$                         &  0.5336($\pm0.0038$)             &  0.5351($\pm0.0016$)     \\
$f$                                   &  $27\,\%$($\pm$4\,\%$$)          &  $29\,\%$($\pm$2\,\%$$) \\
$\theta(^{\circ})$                    &  155.6($\pm0.6$)                 &  153.7($\pm1.1$)   \\
$\psi(^{\circ})$                      &  240.5($\pm2.7$)                 &  278.6($\pm0.5$)  \\
$r$(rad)                              &  0.53(fixed)                     &  0.53(fixed)   \\
$T_f$                                 &  0.73(fixed)                     &  0.73(fixed)  \\
$\Sigma{\omega(O-C)^2}$               &  0.00256                         &  0.00260            \\
\hline
\hline
\end{tabular}
\end{center}
\end{table}

\section[]{DISCUSSION AND CONCLUSION}

According to the photometric solutions in Table \ref{WD_results}, the light curves observed in 1971 and 2018 provide almost consistent results. The two component stars and geometric structure of V752 Cen were quite stable over the past forty-eight years. The solutions show that V752 Cen is a W-subtype contact binary sytem. The star eclipsed at primary minimum is hotter but less massive than the star eclipsed at secondary minimum. The orbital inclination is $i(^{\circ}) = 82.07$, and the totally eclipsed characteristic implies that the determined parameters are very reliable. Considering the mass function obtained by \citet{1974AJ.....79..391S}: $(M_1+M_2)sin^3i = 1.648\pm0.089M_\odot$, the absolute parameters of the two component stars are calculated and listed in Table \ref{absolute}. The orbital semi-major axis is calculated to be $a = 2.59(\pm0.05)R_\odot$. It should be mentioned that the mass ratio of V752 Cen was determined to be $q = 0.31$ from spectroscopic observations \citep{1974AJ.....79..391S}, which was similar to our value (1/3.35 = 0.30).

\begin{table}
\caption{Absolute parameters of components in V752 Cen}\label{absolute}
\begin{center}
\small
\begin{tabular}{lllllllll}
\hline
Parameters                        &Primary                         & Secondary          \\
\hline
$M$                               & $0.39(\pm0.02)M_\odot$         & $1.31(\pm0.07)M_\odot$         \\
$R$                               & $0.77(\pm0.01)R_\odot$         & $1.30(\pm0.02)R_\odot$         \\
$L$                               & $0.75(\pm0.03)L_\odot$         & $2.00(\pm0.07)L_\odot$         \\
\hline
\end{tabular}
\end{center}
\end{table}

V752 Cen is a triple-lined spectroscopic quadruple system, and the tertiary and fourth components may make up a far away binary system \citep{2009ASPC..404..199S}. We have also tried to set third light ($l_3$) as a free parameter while modeling the light curve with the W - D program. However, the solution is not convergent. It means that the potential binary orbiting around V752 Cen should be late-type stars and don't contribute any light to the light curves of V752 Cen. The period analysis of V752 Cen reveals that its period was constant from 1970 to 2000, which means there was no mass transfer between the two component stars during that time interval. Then, its period changed suddenly around the year 2000. However, photometric analysis implies that the component stars in central binary system are very stable. Thus, the sudden period change may have been caused by the interaction between V752 Cen and the potential binary that is orbiting it. The period of V752 Cen has been increasing continuously at a rate of $dP/dt=+5.05\times{10^{-7}}day\cdot year^{-1}$ ($+0.044 s\cdot year^{-1}$) after that sudden period change happened. The period change is almost the same as that determined by \citet{2015A&A...578A.136L}, which was $+0.04379 s\cdot year^{-1}$. It can be explained by mass transfer from the less massive component star to the more massive one, and the mass transfer rate is estimated to be $\frac{dM_{2}}{dt}=2.52\times{10^{-7}}M_\odot/year$.

Both spectroscopic and photometric observations have indicated that many contact binaries are accompanied by at least one additional companion \citep{2006AJ....131.2986P,2013ApJ...768...33R}. Therefore, hierarchical contact binary systems are probably common. What is more important is that all short
period M-dwarf contact binaries may have additional companions since the time-scale of angular momentum loss for the components is too long \citep{2011AcA....61..139S}. \citet{2014ApJ...793..137N} claim that the eccentric Kozai-Lidov mechanism may remove angular momentum from central binaries and result in tidal tightening of inner binaries in triple star systems. However, no observational evidence on dynamic interactions between inner binaries and additional companions was reported. In that respect, V752 Cen is a very interesting hierarchical contact binary system with significant research value. The research on its light curves and period variations suggests that dynamic interactions may have happened between the inner binary and companion stars orbiting it. However, more observations and analyses on hierarchical contact binary systems are still needed to investigate the dynamic mechanism between the central binary and its companion stars in detail.

\section*{Acknowledgements}

We appreciate the valuable comments and suggestions from the anonymous referee. We are grateful to Professor Wayne Orchiston for improving the manuscript. This research was supported by the National Natural Science Foundation of China (Grant No. 11703080 and 11703082), the Joint Research Fund in Astronomy (Grant No. U1931101) under cooperative agreement between the National Natural Science Foundation of China and Chinese Academy of Sciences, and the Yunnan Natural Science Foundation (Grant No. 2018FB006 and 2016FB004). It was part of the research activities at the National Astronomical Research Institute of Thailand (Public Organization). This work has made use of data from the European Space Agency (ESA) mission {\it Gaia}, processed by the {\it Gaia} Data Processing and Analysis Consortium (DPAC). Funding for the DPAC has been provided by national institutions, in particular the institutions participating in the {\it Gaia} Multilateral Agreement. This paper makes use of data from the DR1 of the WASP data as provided by the WASP consortium, and the computing and storage facilities at the CERIT Scientific Cloud, reg. no. CZ.1.05/3.2.00/08.0144 which is operated by Masaryk University, Czech Republic.

\label{lastpage}
\end{document}